\documentstyle[aps]{revtex}

\begin{document}
\title{Noise induced inflation}
\author{{\footnotesize Esteban Calzetta }}
\address{$\dagger $Instituto de Astronom\'{\i}a y F\'{\i}sica del\\
Espacio (IAFE) and Departamento de F\'{\i}sica,\\
Universidad de Buenos Aires,Argentina}
\maketitle

\begin{abstract}
We consider a closed Friedmann-Robertson-Walker Universe driven by the back
reaction from a massless, non-conformally coupled quantum scalar field. We
show that the back-reaction of the quantum field is able to drive the
cosmological scale factor over the barrier of the classical potential so
that if the universe starts near zero scale factor (initial singularity) it
can make the transition to an exponentially expanding de Sitter phase, with
a probability comparable to that from quantum tunneling processes. The
emphasis throughout is on the stochastic nature of back reaction, which
comes from the quantum fluctuations of the fundamental fields.
\end{abstract}

\section{Introduction}

In this talk, based on work done in collaboration with Enric Verdaguer \cite
{paper}, we shall discuss a model of the Early Universe where a spatially
closed, Friedmann - Robertson - Walker Universe avoids recollapse and
launchs into inflationary expansion due to the effects of the back reaction
of quantized matter fields. Central to the argument is the fact that this
back reaction is both memory dependent and to some extent random, due to
quantum fluctuations of the fundamental fields. The stress in particle
creation and noise is the main novelty of our approach with respect to the
by now large corpus of semiclassical cosmology \cite{BD82}.

To the best of our knowledge, this work is the most ellaborate application
to date of the basic idea that quantum fluctuations of fundamental fields
act on the geometry of the Universe as a stochastic energy -momentum tensor,
put forward by several researchers \cite
{CH94,CV94,HuSin,CV96,CV97,LM97,CCV97}. We shall pass briefly on the
technical details, which are contained in the original work, to concentrate
on the physical ideas. In next Section, we give a general discussion of why
and how stochastic terms ought to be included in Einstein equations; the
following Section describes the model and the development of its solution,
and we conclude with some brief Final Remarks.

\section{The semiclassical approximation: particle creation and noise}

Ever since the development of quantum mechanics and relativity theory early
in this century, their final unification has been one of the most sought
after prizes of theoretical physics. Moreover, as Hubble's observations
first, and then the discovery of the cosmic microwave radiation, has taught
us that the structure of our Universe is determined by events at the very
beginning of its evolution, this search for unification acquired more than
academic interest, since these early stages presumably demand a quantum
description. However, in spite of some progress, specially in the sixties,
the goal of a quantum theory of gravity seems now as elusive as ever \cite
{QG}.

In the meantime, it has been realized that even without a full theory it was
possible to find answers to most pressing questions, at least from the
cosmological and astrophysical viewpoints. Basically the theoretical
expectation is that quantum gravitational effects will become relevant at
the Planck scale (of $10^{19}GeV$ in natural units), while quantum effects
associated to matter are conspicuous at the scales associated to the masses
of those particles, orders of magnitude below that. Therefore, it makes
sense to develop models where the geometry of the Universe (or, say, the
space surrounding a compact object) is treated in the terms of Einstein's
theory (which is to say, classically) while matter is quantized \cite{BD82}.
As shown by Parker \cite{Parker}, even these restricted models leave ample
room for exotic behavior, probably its most spectacular manifestation being
Hawking radiation \cite{blackholes,CGHS}.

Now, there being no question that quantum fluctuations {\it weigh}, as shown
by the Casimir effect, its presence should affect the evolution of the said
geometry, on equal footing with other forms of matter, gravity included. So,
although models of quantum test fields on given backgrounds have been
immensely useful to clarify the formal aspects of the theory, the real goal
is to develop self consistent models, where the feedback between the quantum
matter and the classical geometry is fully accounted for. This raises the
issue of how the quantum matter affects the geometry.

The simplest and earliest answer to this question is that it is possible
that while the field amplitudes associated to matter are to be described by
operators in some functional space, other observables such as energy -
momentum, which are composites of field operators, may develop a condensate
or c-number part. In the simplest case, this condensate may be computed as
the expectation value of the corresponding composite operator in an adequate
quantum state, and this expectation value is to be included as the quantum
matter contribution to Einstein's equations \cite{R63}.

By now, there is a substantial body of work showing that this approach can
be made mathematically consistent \cite{BD82}. However, there are also
serious doubts concerning whether it is {\it physically }satisfactory \cite
{fordkuo,hunich}. One can formulate these doubts with any degree of
sofistication, but the basic argument goes as follows.

In a typical self - consistent evolution problem, the identification of the
energy - momentum tensor of the matter fields involves a normal ordering or
subtraction procedure, to take care of the ever present divergences of
quantum field theory. Suppose this normal ordering amounts to substract the
energy - momentum defined with respect to some local vacuum state, the most
sensible choice being a local adiabatic vacuum of some order. Suppose we let
the model evolve from some time $t_i$ to some final time $t_f$. In almost
every interesting case, the evolving geometry will mix the positive and
negative frequency components of the field operator, so that the creation
and destruction operators of the adiabatic model at time $t_f$ will be
related to that at time $t_i$ by a Bogolubov transformation. This
transformation is characterized by two complex parameters, $\alpha $ and $%
\beta $, with $\left| \alpha \right| ^2-\left| \beta \right| ^2=1.$ If at
time $t_i$ a given one particle state was occupied by $n$ adiabatic
particles, then at time $t_f$ we find, in average, $n^{\prime }=\left| \beta
\right| ^2+n\left| \alpha \right| ^2$ particles there (we assume Bose
statistics, for concreteness). This average is reflected in the evolution of
the mean energy -momentum. But the state at times $t_f$ will not, in
general, be a state with a well defined occupation number (certainly not, if
the state at time $t_i$ was), and when we look at the dispersion in particle
number, we see that unless $\left| \beta \right| ^2\ll n$, in which case we
did not need to bother with self consistency, the average fluctuation in
particle number is of the order of the mean occupation number itself. So the
average energy momentum of the created particles is only vaguely related to
what might have happened ''on the spot''.

The issue then arises of how to introduce these fluctuations in an actual
model. Maybe the final answer is that nothing short of full quantum gravity
is truly satisfactory, but if one wishes to retain the classical character
of the metric, then it seems that the only possibility is to add, to the
mean energy - momentum of the fields, a stochastic component, which would
represent the leading effects of the quantum fluctuations.

J. Halliwell and others \cite{DH98,H98} have developed a picture which shows
that this classical, stochastic energy momentum at least makes sense. In
this approach, the classical geometry is seen as an apparatus which measures
the energy - momentum of the quantum matter fields, and reacts to the
measured value. It is then seen that the results are c - numbers, but they
progression in time does not follow deterministic laws. It is natural to
extract the deterministic, mean evolution, and call the remainder random
noise. In the limit where the measurements are inaccurate enough, and
repeated often enough, we may assume that the evolution of the fluctuations
will not be greatly perturbed by the measurement process, and so they may be
computed from ordinary quantum field theoretical rules.

\subsection{The closed time path effective action}

In this limit, therefore, a new paradigm appears, derived from the work of
Feynman and Vernon \cite{Feynman}. We now regard the geometry as an {\it %
open system} evolving in the environment provided by the matter quantum
fluctuations \cite{opensys}. Since the detailed evolution of the later is
deemed irrelevant, our only concern is to estimate the {\it influence action}%
, namely, the modification to the gravitational action due to the influence
of the environment. In the limit where the geometry is actually taken to be
classical, the task becomes identical to that of computing the so-called
Schwinger - Keldysh effective action \cite{Schwinger}. We shall disregard
the somewhat technical distinctions between these two objects, regarding
them as the same; in a nutshell, the Schwinger - Keldish or closed time path
(CTP) effective action (EA) is the influence action evaluated over an
infinite time range, which actually takes care of some difficulties in the
evaluation of the influence action over finite lapses.

The CTPEA is a truly remarkable object, which achieves the miracle of
providing a well defined variational method to derive causal but non local
in time equations of motion. It is not hard to see where the difficulty
lies. Suppose you have a system described by some variable $\phi \left(
t\right) $, and write for it an action functional $S\left[ \phi \right] $
whose variation yields the equations of motion $\delta S/\delta \phi \left(
t\right) =0$. If the equations are causal, then $\delta ^2S/\delta \phi
\left( t\right) \delta \phi \left( t^{\prime }\right) =0$ whenever $%
t^{\prime }>t$. Since second derivatives commute, the second derivative
actually vanishes if only $t^{\prime }\neq t$, and the action must be
necessarily local in time. Since all dissipative effects are physically
limiting cases of non local time interactions (when the response time of the
bath is much shorter than the characteristic time of the system), it follows
more generally that there is no variational principle for dissipative,
causal evolutions.

The CTPEA achieves the impossible by adding to every degree of freedom $\phi
^{+}\left( t\right) $ a mirror degree of freedom $\phi ^{-}\left( t\right) $%
, so that the CTPEA $\Gamma =\Gamma \left[ \phi ^{+},\phi ^{-}\right] $ and
the equations of motion are $\delta \Gamma /\delta \phi ^{+}=0$. The right
count of degrees of freedom is restored by imposing, {\it after the
variation has been taken}, the constraint $\phi ^{+}=\phi ^{-}=\phi $, the
physical degree of freedom. Causality only demands

\begin{equation}
\frac{\delta ^2\Gamma }{\delta \phi ^{+}\left( t\right) \delta \phi
^{+}\left( t^{\prime }\right) }+\frac{\delta ^2\Gamma }{\delta \phi
^{+}\left( t\right) \delta \phi ^{-}\left( t^{\prime }\right) }%
=0\;when\;t<t^{\prime }  \label{causality}
\end{equation}

For example, suppose that the solution to the equations of motion is just $%
\phi =0$, and we seek the dynamics of small fluctuations. The only quadratic
action compatible with the causality constraint has the form

\begin{equation}
\Gamma =\frac 12\int dtdt^{\prime }\;\left\{ \left[ \phi \right] \left(
t\right) {\cal D}\left( t,t^{\prime }\right) \left\{ \phi \right\} \left(
t^{\prime }\right) +\left[ \phi \right] \left( t\right) {\cal N}\left(
t,t^{\prime }\right) \left[ \phi \right] \left( t^{\prime }\right) \right\}
\label{quadratic}
\end{equation}

where $\left[ \phi \right] =\phi ^{+}-\phi ^{-}$; $\left\{ \phi \right\}
=\phi ^{+}+\phi ^{-}$, and ${\cal D}\left( t,t^{\prime }\right) =0$ if $%
t<t^{\prime }$. The equations of motion are

\begin{equation}
\int dt^{\prime }\;{\cal D}\left( t,t^{\prime }\right) \phi \left( t^{\prime
}\right) =0  \label{equations}
\end{equation}

and we see that there is no obstacle to causality, with no further
restrictions on locality.

To actually compute the CTPEA, Schwinger observed that the mean value of the
field could be obtained from a generating functional

\begin{equation}
Z\left[ J^{+},J^{-}\right] =e^{iW\left[ J^{+},J^{-}\right] }=\left\langle 
\tilde T\left( e^{-i\int J^{-}\Phi }\right) T\left( e^{i\int J^{+}\Phi
}\right) \right\rangle  \label{genfun}
\end{equation}

where $\Phi $ is the Heisenberg field operator, the expectation value is
taken with respect to the corresponding quantum state, and $T$, $\tilde T$
stand for temporal and antitemporal ordering, respectively. Indeed, if we
define

\begin{equation}
\phi ^{\pm }=\pm \frac{\delta W}{\delta J^{\pm }}  \label{meanfields}
\end{equation}

Then in the limit $J^{\pm }\rightarrow 0$, we obtain $\phi ^{+}=\phi
^{-}=\left\langle \Phi \right\rangle .$ This suggests defining $\Gamma $ as
the Legendre transform of $W$

\begin{equation}
\Gamma \left[ \phi ^{+},\phi ^{-}\right] =W\left[ J^{+},J^{-}\right] -\int
\left( J^{+}\phi ^{+}-J^{-}\phi ^{-}\right)  \label{ctpea}
\end{equation}

So in general we obtain

\[
\frac{\delta \Gamma }{\delta \phi ^{\pm }}=\mp J^{\pm } 
\]

and for the physical mean field

\begin{equation}
\left. \frac{\delta \Gamma }{\delta \phi ^{\pm }}\right| _{\phi ^{+}=\phi
^{-}=\left\langle \Phi \right\rangle }=0  \label{ctpeqs}
\end{equation}

as required. Observe that on top of Eq. (\ref{causality}), the quantum CTPEA
obeys $\Gamma \left[ \phi ^{+},\phi ^{-}\right] =-\Gamma \left[ \phi
^{-},\phi ^{+}\right] ^{*}$, so that the kernel ${\cal D}$ must be real (and
so are the equations of motion) and ${\cal N}$ is pure imaginary (${\cal N}%
=iN).$

The CTPEA admits a functional representation derived from the usual
background field methods \cite{effaction}

\begin{equation}
e^{i\Gamma \left[ \phi ^{+},\phi ^{-}\right] }=\int D\varphi ^{+}D\varphi
^{-}\;\exp i\left[ S\left( \varphi ^{+}\right) -S\left( \varphi ^{-}\right) -%
\frac{\delta \Gamma }{\delta \phi ^{+}}\left( \varphi ^{+}-\phi ^{+}\right) +%
\frac{\delta \Gamma }{\delta \phi ^{-}}\left( \varphi ^{-}-\phi ^{-}\right)
\right]  \label{bfm}
\end{equation}

The variables of integration must coincide at some very large time $T_\infty 
$ in the future, and we have included the information on the quantum state
in the integration measure. In our case, we have both system $\left( \varphi
\right) $ and environment $\left( \psi \right) $ degrees of freedom, but we
associate a mean field only to the former. The classical action $S$ will be
the sum of the system action $S_s$, the environment action $S_e$, and the
interaction term $S_i$. In the semiclassical limit, we neglect the deviation
from the mean of the system variables, and the expression for the CTPEA\
simplifies to

\begin{equation}
e^{i\Gamma \left[ \phi ^{+},\phi ^{-}\right] }=e^{i\left[ S_s\left( \phi
^{+}\right) -S_s\left( \phi ^{-}\right) \right] }\int D\psi ^{+}D\psi
^{-}\;\exp i\left[ S_e\left( \psi ^{+}\right) +S_i\left( \phi ^{+},\psi
^{+}\right) -S_e\left( \psi ^{-}\right) -S_i\left( \phi ^{-},\psi
^{-}\right) \right]  \label{ctpia}
\end{equation}

where, again, the integration is nontrivial thanks to the integration
measure and the future boundary conditions.

Eq. (\ref{ctpia}) provides a well defined recipe for the CTPEA, and from
thence finding the effective equations of motion is only a matter of
computing power. The question arises of when the solutions to the
semiclassical equations are relevant to the description of our experience.
Mostly what we wish to do is to compute expectation values of system
observables. In so far as these observables do not involve environmental
variables, their expectation values will admit representations such as

\begin{equation}
{\cal A}=\int D\phi ^{+}D\phi ^{-}\;A\left[ \phi ^{+},\phi ^{-}\right]
e^{i\Gamma \left[ \phi ^{+},\phi ^{-}\right] }  \label{expval}
\end{equation}

which under the saddle point approximation reduces to

\begin{equation}
{\cal A}=A\left[ \phi ,\phi \right]  \label{saddle}
\end{equation}

$\phi $ being the solution to the mean field equations. In other words, the
semiclassical equations will be useful if our only interest is to compute
expectation values for system observables, such as the integral Eq. (\ref
{expval}) may be evaluated by saddle point methods.

The CTPEA is a powerful method to derive equations of motion for open (or
effectively open) systems, which are guaranteed to be both real and causal 
\cite{CTP,CH87}. It is possibly the most powerful method at hand to study
nonequilibrium evolution of quantum fields, specially if combined with more
sophisticated resummation methods, which allow us to keep track of higher
Schwinger functions along with the mean field \cite{CH88,BDV}. But two
things ought to send an alarm signal. First, if we only care about the
equations of motion, we are only using a small part of the information
encoded in the CTPEA; for example, the kernel ${\cal N}$ is irrelevant to
the linearized equations (\ref{equations}). Second, there is no noise in the
mean field equations. We must stress that this is not only unseeming, but,
insofar as the equations of motion are dissipative, it is actually {\it %
wrong, }as it violates the necessary balance between fluctuations and
dissipation \cite{FDT}.

\subsection{Where is the noise?}

To the best of our knowledge, it was Feynman who pointed out that the
untapped terms in the CTPEA contained the information about noise \cite
{Feynman}. For simplicity, assume the CTPEA has the quadratic form Eq.(\ref
{quadratic}). Then we have the identity

\begin{equation}
e^{i\Gamma \left[ \phi ^{+},\phi ^{-}\right] }=\int Dj\;\rho \left[ j\right]
\exp \left\{ \frac i2\int dtdt^{\prime }\;\left[ \phi \right] \left(
t\right) {\cal D}\left( t,t^{\prime }\right) \left\{ \phi \right\} \left(
t^{\prime }\right) +i\int dt\;j\left( t\right) \left[ \phi \right] \left(
t\right) \right\}   \label{gauss}
\end{equation}

with

\begin{equation}
\left\langle j\left( t\right) j\left( t^{\prime }\right) \right\rangle
\equiv \int Dj\;\rho \left[ j\right] j\left( t\right) j\left( t^{\prime
}\right) =N\left( t,t^{\prime }\right)  \label{noise}
\end{equation}

Now we can write the average Eq. (\ref{expval})

\begin{equation}
{\cal A}=\int Dj\;\rho \left[ j\right] \int D\phi ^{+}D\phi ^{-}\;A\left[
\phi ^{+},\phi ^{-}\right] \exp \left\{ \frac i2\int dtdt^{\prime }\;\left[
\phi \right] \left( t\right) {\cal D}\left( t,t^{\prime }\right) \left\{
\phi \right\} \left( t^{\prime }\right) +\int dt\;j\left( t\right) \left[
\phi \right] \left( t\right) \right\}  \label{noiseev}
\end{equation}

And use saddle point evaluation in the inner integral

\begin{equation}
{\cal A}=\int Dj\;\rho \left[ j\right] A\left[ \phi _j,\phi _j\right]
\label{saddlenoise}
\end{equation}

where $\phi _j$ is the solution to

\begin{equation}
\int dt^{\prime }\;{\cal D}\left( t,t^{\prime }\right) \phi \left( t^{\prime
}\right) =-j(t)  \label{langevin}
\end{equation}

In this way, we have transformed the average of the observable ${\cal A}$
over the quantum fluctuations of the environment into the ensemble average
over the realizations of the ''noise'' $j(t)$, at the same time upgrading
the semiclassical equations to the Langevin equations Eq. (\ref{langevin}).
As expected, the relevant information on the noise (its correlation
function), is given by the ''useless'' part of the CTPEA, namely the kernel $%
N=-i{\cal N}$.

Of course, eq. (\ref{gauss}) is not the only way to decompose the CTPEA into
partial integrations. The point is that this particular decomposition makes
physical sense. To see this, assume the interaction term in the action takes
the particular form

\begin{equation}
S_i=\int dt\;\Xi \left[ \psi \right] \phi  \label{linear}
\end{equation}

with $\left\langle \Xi \right\rangle =0$ when $\phi =0$ but otherwise
arbitrary. Then \cite{CH94}

\begin{equation}
N(t,t^{\prime })=\frac 12\left\langle \left\{ \Xi (t),\Xi (t^{\prime
})\right\} \right\rangle  \label{anticom}
\end{equation}

the expectation value being computed at $\phi =0$. Indeed, the Heisenberg
equation for this model is

\begin{equation}
\frac{\delta S_s}{\delta \phi }=-\Xi \left[ \psi \right]  \label{heisen}
\end{equation}

We assume that the Heisenberg operator for the system variable is close to a
c-number. Also, in the presence of a non zero background $\phi $, the
operator $\Xi $ will generally develop a nonzero expectation value $%
\left\langle \Xi \right\rangle _\phi $. Subtracting this, we get

\begin{equation}
\frac{\delta S_s}{\delta \phi }+\left\langle \Xi \right\rangle _\phi
=-\left( \Xi -\left\langle \Xi \right\rangle _\phi \right)  \label{noisehei}
\end{equation}

The CTPEA, if we forget about noise, leads to the same equation with no
right hand side. As discussed above, this is not acceptable. So the question
is, what is the sensible way of replacing the q-number operator in the right
hand side of Eq. (\ref{noisehei}) by a c-number stochastic source. Of
course, some loss of information (specially concerning quantum coherence) is
unavoidable, and in particular cases this may invalidate the whole
procedure. But whenever quantum coherence is not the main concern, a
Gaussian source with self correlation as in Eq. (\ref{anticom}) (we again
neglect $\left\langle \Xi \right\rangle _\phi ,$ which vanishes at $\phi =0$%
, since we assume small fluctuations, but this is not essential) is the time
honored answer, and indeed the only answer compatible with the fluctuation
dissipation theorem \cite{landau}.

\section{Noise Induced Inflation: when noise matters}

As we have seen in the previous Section, the CTPEA provides a systematic
framework in which to study semiclassical evolution, taking into account at
least the leading effects due to quantum fluctuations of matter fields.
However, putting this framework into work is by no means a simple task.
Indeed, after the observation that the CTPEA provides a simple way of
deriving Einstein - Langevin equations \cite{CH94}, it was realized that
actually carrying out the derivation is a research project in itself \cite
{CV94,HuSin,CV96,CV97,LM97}, to say nothing of solving those equations once
derived \cite{CCV97}. So it is natural to wonder if noise makes such a
difference as to justify this trouble.

The basic problem is that, while it would be easy to find problems where the
noise level is huge, this same noisyness would lead to the suspicion that
the whole semiclassical approach is breaking down. The real challenge is to
find a problem where the semiclassical approximation is reliable, and still
noise makes a difference. Indeed, in the case of conventional, noiseless
semiclassical theory, such a problem is Hawking evaporation of large black
holes: a weak effect, which puts in no jeopardy the validity of the
semiclassical approximation, but whose result is utterly impossible in terms
of the classical theory alone.

In this talk, I will report on one such problem, a cyclic Universe, provided
with a cosmological constant but prevented from inflating by the potential
barrier from its own spatial curvature. In each cycle, the semiclassical
effects induce a transition form one classical orbit to another; the change
is small for each cycle, but overall it offers an escape route with no
classical analog. Our goal is to compute the average escape probability due
to semiclassical effects.

This problem has an important precedent, the calculation of the tunneling
amplitude due to Vilenkin \cite{V8284}. It is important to realize the
similarities and differences between these two approaches. Vilenkin's
calculation was fully quantum gravitational, but it contemplated only the
effects of the gravitational field. It was tacitly assumed that, if any
matter fields were present, they would at most affect the prefactor of the
exponentially suppressed tunneling probability \cite{Coleman,Affleck,Garriga
1994,Barvinsky,Caldeira 1983,Rubakov 1984}. Our calculation is only
semiclassical, but we put the stress precisely on the effects of the matter
fields. From the point of view of the usual instanton approach, we could say
ours is a highly nonperturbative evaluation of the prefactor, since we go
well beyond the test field - one loop approximation. The result is that the
escape probability due to the fluctuations in matter fields is at least as
large as the tunneling one, suggesting that in Nature both must be taken
into account. We shall not discuss subsequent developments related to
Vilenkin 's proposal \cite{Jerusalem}.

Since the calculation of the escape probability due to semiclassical effects
is discussed in some detail elsewhere \cite{paper}, here I shall only give a
general discussion of the several steps involved, the peculiar difficulties
of each one, and how they could be overcome.

\subsection{The model}

Our model is based on a spatially closed, homogeneous Friedmann - Robertson
- Walker (FRW) model, with a metric

\begin{equation}
ds^2=a^2(t)\left( -dt^2+\tilde g_{ij}(x^k)dx^idx^j\right) ,\ \ \ \
i,j,k=1,...,n-1,  \label{1}
\end{equation}
where $a(t)$ is the cosmological scale factor, $t$ is the conformal time,
and $\tilde g_{ij}(x^k)$ is the metric of an $(n-1)$-sphere of unit radius.
Since we will use dimensional regularization we work, for the time being, in 
$n$-dimensions. Matter is described by a quantum scalar field $\Phi (x^\mu )$%
, where the Greek indices run from $0$ to $n-1$. The classical action for
this scalar field in the spacetime background described by the above metric
is 
\begin{equation}
S_m=-\int dx^n\sqrt{-g}\left[ g^{\mu \nu }\partial _\mu \Phi ^{*}\partial
_\nu \Phi +\left( {\frac{n-2}{4(n-1)}}+\nu \right) R\Phi ^{*}\Phi \right] ,
\label{2}
\end{equation}
where $g_{00}=a^2$, $g_{0i}=0$, $g_{ij}=a^2\tilde g_{ij}$, $g$ is the metric
determinant, $\nu $ is a dimensionless parameter coupling the field to the
spacetime curvature ($\nu =0$ corresponds to conformal coupling), $R$ is the
curvature scalar which is given by 
\begin{equation}
R=2(n-1){\frac{\ddot a}{a^3}}+(n-1)(n-4){\frac{\dot a^2}{a^4}}+(n-1)(n-2){%
\frac 1{a^2}},  \label{3}
\end{equation}
where an over dot means derivative with respect to conformal time $t$. Let
us now introduce a conformally related field $\Psi $ 
\begin{equation}
\Psi =\Phi a^{{\frac{n-2}2}},  \label{4}
\end{equation}
the time dependent function $U(t)$ 
\begin{equation}
U(t)=-\nu a^2(t)R(t),  \label{8}
\end{equation}
and the d'Alambertian $\Box =-\partial _t^2+\Delta ^{(n-1)}$ of the static
metric $\tilde ds^2=a^{-2}ds^2$. The action may be written as, 
\begin{equation}
S_m=\int dtdx^1\dots dx^{n-1}\sqrt{\tilde g}\left[ \Psi ^{*}\Box \Psi -{%
\frac{(n-2)^2}4}\Psi ^{*}\Psi +U(t)\Psi ^{*}\Psi \right] .  \label{6}
\end{equation}

When $\nu=0$ this is the action of a scalar field $\Psi$ in a background of
constant curvature. The quantization of this field in that background is
trivial in the sense that a unique natural vacuum may be introduced, the
``in'' and ``out'' vacuum coincide and there is no particle creation \cite
{BD82}. This vacuum is, of course, conformally related to the physical
vacuum, see (\ref{4}). The time dependent function $U(t)$ will be considered
as an interaction term and will be treated perturbativelly. Thus we will
make perturbation theory with the parameter $\nu$ which we will assume small.

To carry on the quantization we will proceed by mode separation expanding $%
\Psi (x^\mu )$ in terms of the $(n-1)$-dimensional spherical harmonics $Y_{%
\vec k}^l(x^i).$ The coefficients $\Psi _{\vec k}^l(t)$ are just functions
of $t$ ($1$-dimensional fields), and for each set $(l,\vec k)$ we may
introduce two real functions $\phi _{\vec k}^l(t)$ and $\tilde \phi _{\vec k%
}^l(t)$ defined by 
\begin{equation}
\Psi _{\vec k}^l\equiv {\frac 1{\sqrt{2}}}\left( \phi _{\vec k}^l+i\tilde 
\phi _{\vec k}^l\right) ,  \label{15}
\end{equation}

The action becomes the sum of the actions of two independent sets formed by
an infinite collection of decoupled time dependent harmonic oscillators 
\begin{equation}
S_m={\frac 12}\int dt\sum_{k=1}^\infty \sum_{\vec k}\left[ \left( \dot \phi
_{\vec k}^l\right) ^2-M_k^2\left( \phi _{\vec k}^l\right) ^2+U(t)\left( \phi
_{\vec k}^l\right) ^2\right] +...\,,  \label{16}
\end{equation}
where the dots stand for an identical action for the real $1$-dimensional
fields $\tilde \phi _{\vec k}^l(t)$.

We will consider, from now on, the action for the $1$-dimensional fields $%
\phi _{\vec k}^l$ only. The field equation for the $1$-dimensional fields $%
\phi _{\vec k}^l(t)$ are, from (\ref{16}), 
\begin{equation}
\ddot \phi _{\vec k}^l+M_k^2\phi _{\vec k}^l=U(t)\phi _{\vec k}^l,
\label{18}
\end{equation}
which in accordance with our previous remarks will be solved perturbatively
on $U(t)$. The solutions of the unperturbed equation can be written as
linear combinations of the normalized positive and negative frequency modes, 
$f_k$ and $f_k^{*}$ respectively, where 
\begin{equation}
f_k(t)={\frac 1{\sqrt{2M_k}}}\exp (-iM_kt).  \label{19}
\end{equation}

\subsection{Closed time path effective action}

We are now in the position to compute the regularized semiclassical CTP
effective action. This involves a careful consideration of the infinities
arising in perturbation theory, but after the dust settles, the result is 
\begin{equation}
\Gamma _{CTP}[a^{\pm }]=S_{g,m}^R[a^{+}]-S_{g,m}^R[a^{-}]+S_{IF}^R[a^{\pm }],
\label{43a}
\end{equation}

where the regularized gravitational and classical matter actions are, 
\begin{equation}
S_{g,m}^R[a]={\frac{2\pi ^2}{l_P^2}}\int dt\,6a^2\left( {\frac{\ddot a}a}%
+1\right) -2\pi ^2\int dt\,a^4\Lambda ^{\star }+{\frac 1{16}}\int
dt\,U_1^2(t)\ln (a\mu _c).  \label{43b}
\end{equation}

and the influence action 
\begin{equation}
S_{IF}^R[a^{\pm }]={\frac 12}\int dtdt^{\prime }\,\Delta U(t)H(t-t^{\prime
})\{U(t^{\prime })\}+{\frac i2}\int dtdt^{\prime }\,\Delta U(t)N(t-t^{\prime
})\Delta U(t^{\prime }),  \label{43c}
\end{equation}

where we have defined 
\begin{equation}
\Delta U=U^{+}-U^{-},\ \ \ \ \ \{U\}=U^{+}+U^{-}.  \label{44b}
\end{equation}

Computing the kernels $H$ and $N$ involves the consideration of Feynman
graphs, where the internal legs represent propagators for a particle in a
closed space. This calculation may be carried out exactly, but the result is
that, unless for orbits with very small amplitude, the effect of spatial
curvature is not really important. It is convenient to compute these kernels
as in a spatially flat FRW Universe with the same radius, which amounts to
consider a continuous, rather than a discrete, spectrum of modes. The result
is

\begin{equation}
N(u)=\int_0^\infty dk\cos 2ku={\frac \pi {16}}\delta (u).  \label{53a}
\end{equation}

\begin{equation}
H(u)=\frac 18{\rm Pf}\left[ \frac{\theta (u)}u\right] +\frac{\gamma +\ln \mu
_c}8\delta (u).  \label{53d}
\end{equation}

The distribution ${\rm Pf}(\theta(u)/u)$ should be understood as follows.
Let $f(u)$ be an arbitrary tempered function, then 
\begin{equation}
\int_{-\infty}^\infty du {\rm Pf}\left[ \frac {\theta (u)}{u}\right] f(u)=
\lim_{\epsilon\rightarrow 0^+}\left( \int_\epsilon^\infty du \frac {f(u)}{u}
+f(0)\ln\epsilon\right).  \label{53e}
\end{equation}

The approximation of substituting the exact kernels by their flat space
counterparts is clearly justified when the radius of the universe is large,
which is when the semiclassical approximation works best.

. The imaginary part of the influence action is known \cite
{CH94,HuSin,CV96,hargell,Gleiser and Ramos 1994,CH95,eft} to give the effect
of a stochastic force on the system, and we can introduce an improved
semiclassical effective action, 
\begin{equation}
S_{eff}[a^{\pm };\xi ]=S_{g,m}^R[a^{+}]-S_{g,m}^R[a^{-}]+{\frac 12}\int
dtdt^{\prime }\,\Delta U(t)H(t-t^{\prime })\{U(t^{\prime })\}+\int dt\xi
(t)\Delta U(t),  \label{45a}
\end{equation}
where $\xi (t)$ is a Gaussian stochastic field defined by the following
statistical averages 
\begin{equation}
\langle \xi (t)\rangle =0,\ \ \ \ \langle \xi (t)\xi (t^{\prime })\rangle
=N(t-t^{\prime }).  \label{45b}
\end{equation}

The kernel $H$ in the effective action gives a non local effect (due to
particle creation), whereas the source $\xi$ gives the reaction of the
environment into the system in terms of a stochastic force.

\subsection{The Einstein - Langevin equation}

The dynamical equation for the scale factor $a(t)$ can now be found from the
effective action (\ref{45a}) in the usual way, that is by functional
derivation with respect to $a^{+}(t)$ and then equating $a^{+}=a^{-}\equiv a$%
. These equations include the back-reaction of the quantum field on the
scale factor; they improve the semiclassical equation by taking into account
the fluctuations of the stress-energy tensor of the quantum field \cite
{fordkuo,hunich}. However, they also lead to the typical non physical
runaway solutions due to the higher order time derivatives involved in the
quantum correction terms.

To avoid such spurious solutions we use the method of order reduction \cite
{Flanagan and Wald 1996}. In this method one assumes that the equations
obtained from the CTPEA are perturbative, the perturbations being the
quantum corrections. To leading order the equation reduces to the classical
equation, which, in terms of scaled variables 
\begin{equation}
b(t)\equiv {\frac{\sqrt{24}\pi }{l_P}}a(t),\ \ \ \ \ \ \Lambda \equiv {\frac{%
l_P^4}{12\pi ^2}}\Lambda ^{\star }.  \label{46}
\end{equation}

reads 
\begin{equation}
\ddot b+b\left( 1-{\frac 16}\Lambda b^2\right) =O(\nu ).  \label{fifty}
\end{equation}

The terms with $\ddot b$ or with higher time derivatives in the quantum
corrections are then substituted using recurrently the classical equation (%
\ref{fifty}). In this form the solutions to the semiclassical equations are
also perturbations of the classical solutions. Thus, by functional
derivation of (\ref{45a}), we can write the stochastic semiclassical
back-reaction equation as 
\begin{equation}
\dot p=-V^{\prime }\left( b\right) -\delta V^{\prime }(b)+F(b,p,t)+J(\xi
,b,p),  \label{one}
\end{equation}
where a prime means a derivative with respect to $b$, and we have introduced 
$p\equiv \dot b$. The classical potential $V(b)$ is 
\begin{equation}
V\left( b\right) =\frac 12b^2-\frac \Lambda {24}b^4,  \label{two}
\end{equation}

An schematic plot of this potential is given in Fig. 1.

The remaining terms in eq. (\ref{one}) represent the quantum corrections.
The first one is purely local  
\begin{equation}
\delta V(b)=-\frac{3\nu ^2\Lambda }4\left[ \frac 12b^2-\frac \Lambda {48}%
b^4-p^2\ln (b\bar \mu )\right] ,  \label{three}
\end{equation}
where we have already implemented order reduction.

From this point on, we shall disregard the local quantum correction to the
potential, $\delta V(b)$. In the region where semiclassical theory is
reliable, this is only a very small correction to the classical potential;
moreover, we are concerned with such phenomena where the semiclassical
behavior is qualitatively different from the classical one, which is not the
case for these corrections.

The term $F(b,p,t)$ involves nonlocal contributions and may be written as, 
\begin{equation}
F(b,p,t)=-\frac{\partial U}{\partial b}I-\frac{d^2}{dt^2}\left( \frac{%
\partial U}{\partial \ddot b}I\right) =6\nu \left\{ \frac{d^2}{dt^2}\frac 1b-%
\frac{\ddot b}{b^2}\right\} I,  \label{51}
\end{equation}

where $I(b,p,t)$ is defined by 
\begin{equation}
I(b,p,t)\equiv \int_{-\infty }^\infty dt^{\prime }H(t-t^{\prime
})U(t^{\prime }).  \label{52}
\end{equation}

and

\begin{equation}
U(t)=-6\nu \left( {\frac{\ddot b}b}+1\right) .  \label{48}
\end{equation}

After order reduction, $U\left( t^{\prime }\right) $ must be evaluated on
the classical orbit with Cauchy data $b\left( t\right) =b$, $p\left(
t\right) =p$, whereby it reduces to $U=-\Lambda \nu b^2$. Observe that, in
fact, this approximation makes the equation of motion local in time, though
non longer Hamiltonian. Finally, the function $J$ is the noise given by 
\[
J\left( \xi ,b\right) =6\nu \left\{ \frac{d^2}{dt^2}\left( \frac \xi b%
\right) -\frac{\ddot b\xi }{b^2}\right\} 
\]
and, after order reduction, by 
\begin{equation}
J(\xi ,b,p)=6\nu \left[ \frac{\ddot \xi }b-\frac{2\dot \xi p}{b^2}+\frac{%
2\xi V^{\prime }\left( b\right) }{b^2}+\frac{2\xi p^2}{b^3}\right] ,
\label{six}
\end{equation}
with $\xi (t)$ defined in (\ref{45b}) in terms of the noise kernel.

\subsection{The classical orbits}

Before continuing, it is convenient to pause and consider the classical
orbits, as described by Eq. (\ref{fifty}). They represent a particle moving
in the one dimensional potential well Eq. (\ref{two}), plotted schematically
in Fig. 1. This evolution preserves the Wheeler - DeWitt operator (energy,
for short)

\begin{equation}
H(b,p)=\frac 12p^2+V(b)  \label{54a}
\end{equation}

Physically, the value of $H$ on a given orbit is the energy density of
radiation present besides the cosmological constant.

Fig. 2 is a schematic representation of the classical phase space. There is
a stable fixed point ({\it nothing}) at $b=p=0$. This is the starting point
of Vilenkin's calculation. There is an unstable fixed point corresponding to

\begin{equation}
H=E_s=\frac 3{2\Lambda };\;b=2\sqrt{E_s};\;p=0  \label{unstable}
\end{equation}

This is an Einstein type static Universe. When $H>E_s$, orbits are free to
expand forever.

For $H<E_s$, we have two types of orbits. Those outside the well contract at
first, until they reach the classical turning point $b_{+}$ and bounce off.
The De Sitter Universe, with $H=0$, belongs to this family. In their final
stages, these orbits are essentiallly identical to the ever expanding ones
(cosmic no hair theorem). Orbits inside the well with $0<H<E_s$ bounce
ethernally between the turning points $\pm b_{-}$ (there is no problem with
a negative radius of the Universe, since only $b^2$ has a physical meaning;
we may also think of $b=0$ as a perfectly reflecting boundary). The actual
location of the turning points is 
\begin{equation}
b_{\pm }^2=4E_s\left[ 1\pm \sqrt{1-\frac E{E_s}}\right] ,  \label{bplus}
\end{equation}

The frequency $\Omega $ of oscillation is $1$ for $H\ll E_s$, and vanishes
as $H\rightarrow E_s$. This limiting value corresponds to two orbits,
exponentially departing from and approaching to the unstable fixed points,
the so-called separatrices.

To fix ideas, let us adopt for $\Lambda $ a value consistent with Grand
Unified scale inflation, which in natural units means $\Lambda \sim 10^{-12}$%
. Then the value of the separatrix energy is very high, $E_s\sim 10^{12}$.
Our problem is to find a way for a typical Universe ($H\sim 1$), trapped
within the well, to climb out of it and inflate. As we shall see, this is
possible thanks to the combination of the diffussive effect of quantum
fluctuations and a runaway instability associated to particle creation.

It ought to be clear that the final state of this evolution will be very
different than in the instanton approach. In the quantum claculation, the
Universe emerges from under the barrier as an empty, $H=0$, De Sitter
Universe. In our calculation, the Universe goes above the barrier, and
emerges with a large amount of radiation $H=E_s$, corresponding to particles
created while inside the well. Physically however the difference is minor,
as this energy gets diluted in a few e-foldings by the inflationary
expansion.

\subsection{From Langevin to Kramers}

Now we want to determine the probability that a universe starting at the
potential well goes over the potential barrier into the inflationary stage.
The magic of the CTPEA has turned an originally quantum problem into a
statistical mechanical one, indeed a classic problem associated to the name
of Kramers \cite{Kramers}. Observe that we are not interested in the
features of solutions associated to peculiar realizations of the noise, but
rather on a noise averaged observable. Therefore, it is convenient to
perform the noise average at the outset, introducing the distribution
function 
\begin{equation}
f\left( b,p,t\right) =\left\langle \delta \left( b\left( t\right) -b\right)
\delta \left( p\left( t\right) -p\right) \right\rangle ,  \label{seven}
\end{equation}
where $b(t)$ and $p(t)$ are solutions of equation (\ref{one}) for a given
realization of $\xi (t)$, $b$ and $p$ are points in the phase space, and the
average is taken both with respect to the initial conditions and to the
history of the noise. After some standard manipulations we arrive at the
so-called Kramers' equation \cite{Sancho} 
\begin{equation}
\frac{\partial f}{\partial t}=\{H,f\}-\frac \partial {\partial p}[F(b,p,t)f]-%
\frac \partial {\partial p}\Phi ,  \label{eight}
\end{equation}
where the curly brackets are Poisson brackets, i.e. 
\[
\{H,f\}=-p(\partial f/\partial b)+V^{\prime }(b)(\partial f/\partial p), 
\]
and 
\begin{equation}
\Phi =-\frac{\pi \nu ^2\Lambda ^2}4b^2\frac{\partial f}{\partial p},
\label{Phifinal}
\end{equation}

This term will be called the diffusion term since it depends on the
stochastic field $\xi (t)$.

We notice that in the absense of a cosmological constant, we get no
diffusion. This makes sense, because in that case the classical trajectories
describe a radiation filled universe. Such universe would have no scalar
curvature, and so it should be insensitive to the value of $\nu $ as well.

\subsection{From Kramers to Fokker-Planck}

In the usual statement of Kramers' problem, the system is described by a
single variable $x$ and obeys a Fokker-Planck equation \cite{Risken 1984}.

\begin{equation}
\frac{\partial f}{\partial t}=\gamma \frac \partial {\partial x}\left[ f%
\frac{\partial F}{\partial x}+T\frac{\partial f}{\partial x}\right]
\label{FP}
\end{equation}

where $T$ is the temperature (which fixes the sign of the diffussion term)
and $F$ is the free energy (rather than a potential). Activation is studied
from the properties of the steady solutions of this equation, and the answer
is the so-called Arrhenius formula

\begin{equation}
P\sim e^{-F_{\max }/T}  \label{arrhenius}
\end{equation}

where $F_{\max }$ is the value at the peak of the free energy barrier.

Our Kramers equation is certainly more involved than Eq. (\ref{FP}), because
it describes other phenomena besides tunneling. Basically, there are three
things going on. Given a generic distribution function $f$, its dynamics
consists mostly on the representative phase space points being dragged along
the classical orbits, with a time scale of the order of a typical period. On
a larger time scale, we have the diffussion process, which makes $f$ evolve
towards a quasi equilibrium, steady solution. Finally, there is activation,
on an even larger time scale.

Since our concern is this third process only, it is convenient to get rid of
the two faster ones. We get rid of classical transport by defining a new
distribution function which counts the number of Universes on a given
classical orbit, rather than on a phase space cell. This new distribution
function does not tell us where in the orbit we are, but we do not need that
to study activation. We achieve this by transforming the problem to action -
angle variables, and averaging over the angles \cite{Clasmech}. Finally, we
get rid of the approach to quasi equilibrium by assuming a steady solution
from the beginning.

The averaged Kramers equation becomes, 
\begin{equation}
\frac{\partial f}{\partial t}=\frac{\pi \nu ^2\Lambda ^2}4\frac \partial {%
\partial J}\left\{ \frac{{\bf D}(J)}\Omega \frac{\partial f}{\partial J}-%
{\bf S}f\right\}  \label{b6}
\end{equation}

where 
\begin{equation}
{\bf D}(J)=\frac 1{2\pi \Omega }\int_0^{2\pi }d\theta b^2p^2,  \label{b2}
\end{equation}
\begin{equation}
{\bf S}(J)=\frac{-1}{4\pi ^2}\int_0^{2\pi /\Omega }dt\left( \frac d{dt}%
b^2(t)\right) {\rm Pf}\int_0^\infty \frac{du}ub^2(t-u).  \label{b11}
\end{equation}
This equation may be written as a continuity equation $\partial _tf+\partial
_JK=0$, where the probability flux $K$ may be identified directly from (\ref
{b6}). We see that, as in Kramers' problem, stationary solutions with
positive flux $K_0$ should satisfy 
\begin{equation}
\frac{{\bf D}(J)}\Omega \frac{\partial f}{\partial J}-{\bf S}(J)f=-\frac 4{%
\pi \nu ^2\Lambda ^2}K_0.  \label{b12}
\end{equation}

From now on it is more convenient to use the energy $E$ as a variable
instead of $J$, where $E=H(J)$. ${\bf D}$ and ${\bf S}$ individually behave
as $E^2$ times a smooth function of $E/E_s$, and their ratio is relatively
slowly varying. At low energy, we find ${\bf D\sim }E^2/2$ and ${\bf S\sim }%
E^2/4$. As we approach the separatrix, ${\bf D\rightarrow }0.96\,E_s^2$ and $%
{\bf S\rightarrow }1.18\,E_s^2$. Meanwhile, the ratio of the two goes from $%
0.5$ to $1.23$. This means that we can write the equation for stationary
distributions as 
\begin{equation}
\frac{\partial f}{\partial E}-\beta \left( E\right) f=-\frac 4{\pi \nu
^2\Lambda ^2g\left( E\right) }\left( \frac{K_0}{E^2}\right) ,  \label{b12b}
\end{equation}
where $\beta $ and $g$ are smooth order one functions. There is a
fundamental difference with respect to Kramers' problem, namely the sign of
the second term in the left hand side. In the cosmological problem, the
effect of nonlocality is to favour diffussion rather than hindering it. We
may understand this as arising from a feedback effect associated with
particle creation (see \cite{Parker}).

\subsection{The activation amplitude}

Fig. (3) is a out of scale, schematic plot of the solution of Eq. (\ref{b12b}%
). For $E\ll 1$, the solution diverges as $1/E$; for $E\geq 1$, and for
twelve decades thereafter, it grows exponentially. Of course, our analysis
doeas not hold beyond the separatrix, but it can be shown that $f$ turns
around there, decaying as a power of $E$ as $E\rightarrow \infty $.

This behavior cannot be extrapolated all the way to zero as it would make $f$
non integrable. However we must notice that neither our treatment (i.e., the
neglect of logarithmic potential corrections) nor semiclassical theory
generally is supposed to be valid arbitrarily close to the singularity. Thus
we shall assume that the pathological behavior of Eq. (\ref{b12b}) near the
origin will be absent in a more complete theory, and apply it only from some
lowest energy $E_\delta \sim 1$ on. There are still 12 orders of magnitude
between $E_\delta $ and $E_s$.

We may now estimate the flux by requesting that the total area below the
distribution function should not exceed unity. Unless the lower cutoff $%
E_\delta $ is very small (it ought to be exponentially small on $E_s$ to
invalidate our argument) the integral is dominated by the peak around $E_s$,
and we obtain 
\begin{equation}
K_0\leq ({\rm prefactor})\exp \left[ -\beta (E_s)E_s\right] .  \label{tuna}
\end{equation}

The prefactor depends on $\Lambda $, $\nu $, $g\left( 1\right) $, $\beta
\left( 1\right) $, $\sigma $ and the details of the peak shape, but only
logarithmically on the cut -off. So we can take this as a {\it bona fide}
prediction of noisy semiclassical theory. Using $E_s=3/(2\Lambda )$, $\beta
\left( E_s\right) =1.23$, we get 
\begin{equation}
K_0\leq ({\rm prefactor})\exp \left( -\frac{1.84}\Lambda \right) .
\label{tunb}
\end{equation}

This is the semiclassical result to be compared against the instanton
calculations \cite{V8284}, which yield

\begin{equation}
P\sim \exp \left( -\frac 8\Lambda \right) .  \label{b26}
\end{equation}

We see that the probability of noise activation is indeed larger than that
of quantum tunneling.

A final pertinent question is, was noise truly necessary? After all, one
could imagine there would be particle creation in each cycle, and the
accumulation of these particles alone would make inflation progressively
easier to achieve. Of course, in the absense of diffussion our argument
would need to be entirely redrawn; for example, it is not longer clear than
an arbitrary $f$ would tend towards an steady solution, or on which
timescales, nor that the steady solution would have an acceptable behavior
beyond the separatrix. However, one could try to see what happens if one
simply kills the diffussion term in Eq. (\ref{b12b}). The equation still
admits a solution, but the divergence at the origin gets worse ($1/E^2$),
and indeed the flux seems to depend solely on the value of the infrared cut
- off, $K_0\sim E_\delta $, which cannot be predicted within the
semiclassical theory. It is only the combination of particle creation and
diffussion which sets up the mechanism by which we get a definite result.

On the other hand, an infinity of cycles is not truly necessary. There is a
similar result in models where the Universe is restricted to a single cycle;
the activation probability is somewhat lower, but still higher than the
tunneling one. We can understand this by analogy to the problem of black
hole formation in a box, where a big hole can form by slow accretion of
smaller holes, or by a sudden, single large fluctuation. The difference is,
of course, that in the black hole case the latter is more likely than the
former \cite{WP}.

\section{Final Remarks}

We have reported on a cosmological process where quantum induced noise and
particle creation combine to yield a behavior notoriously different than
expected from classical theory, or even conventional, deterministic
semiclassical gravity. The strenght of the effect is indeed comparable to a
purely quantum calculation, which shows by the side that treating matter as
test fields in quantum gravity may not be justified. We believe this work is
meaningful in at least three different levels:

a) of course, our results are most important as an step forward in the
development of stochastic, semiclassical cosmological models. By now, the
mathematical and physical basis of such models is rather well understood,
but the development of actual models and the gathering of hard predictions
is lagging behind. Our calculation has demanded the application of a number
of techniques which are not common tools of the trade in cosmology, and
could serve as an example for future projects.

b) the relevance to cosmology may seem minor, since it does not seem likely
that our Universe be spatially closed. However, the situation changes if the
original question is rephrased as: is it possible that a horizon size,
overdense region in the early Universe, with a homogeneous but subplanckian
value of the inflaton field, may avoid collapse and inflate? Clasically, the
answer is no, and this negative result may well be the bane of inflationary
models \cite{Vacha}. Our results suggest that semiclassically things may
turn around.

c) finally, it has been observed that noise and dissipation are generic to
all effective theories \cite{eft}. So we must expect that similar results
will be found in the analysis of nucleation phenomena in other effective
theories as well, specially in quantum field theories \cite{GdV}, and in out
of equilibrium situations. Indeed, an approach such as ours seems to be the
only way of analyzing tunneling in situations where the environment changes
on timescales comparable to the time it takes to nucleate a bubble, an
essentially virgin field right now.

We continue our research on all these levels, and hope to report soon on new
results.

\begin{center}
{\normalsize {\bf ACKNOWLEDGMENTS}}
\end{center}

The work reported in this talk has been made in collaboration with Enric
Verdaguer \cite{paper}; the spin on the results and the mistakes, if any,
are all mine. Over time, this collaboration has involved a number of younger
researchers, which have been most effective in keeping us on our toes; it is
a very special pleasure to thank Sonia Gonorazky, Antonio Campos, Albert
Roura and the rest of them.

Our collaboration is part of an effort to develop a new way of doing Early
Universe cosmology, which involves several other groups. Bei-lok Hu is very
much the center of this network, and a constant source of inspiration and
encouragement for us all. We also acknowledge constant exchanges with Diego
Mazzitelli, and among the younger ones, Andrew Matacz, Alpan Raval, Charis
Anastopoulos, Stephen Ramsey, Greg Stephens, Nicholas Phillips, Fernando
Lombardo and Diego Dalvit.

While not directly involved, Ted Jacobson, Jaume Garriga and Juan Pablo Paz
have contributed to this program more than they think. In particular, Ted
Jacobson suggested the title of this talk.

In preparing the manuscript for the proceedings, I freely took advantage of
the valuable comments from the audience on delivery. We are particularly
grateful to Alex Vilenkin, Larry Ford, Jonathan Halliwell and Pasqual
Nardone. Finally, I heartfully thank the organizers for puttting the extra
effort of bringing a speaker from faraway Buenos Aires, and for the
wonderfull atmosphere of the whole meeting.

This work has been partially supported by the European project CI1-CT94-0004
and by the CICYT contracts AEN95-0590, Universidad de Buenos Aires, CONICET
and Fundaci\'on Antorchas.

\section{Figure Captions:}

\subsection{Fig 1.}

A schematic plot of the classical potential; it vanishes as $b^2$ when $%
b\rightarrow 0$, it has a maximum, and decreases without bound for larger
Universes. Classical evolution preserves the Wheeler - DeWitt operator $%
H=p^2/2+V(b)$. If $H$ exceeds the maximum of the potential, the
corresponding orbit either expands forever or collapses to the singularity.
For lower $H$, the classical orbit  bounces off the outer classical turning
point. For positive $H$ below the maximum of the potential, we have periodic
bounded orbits representing an eternal cyclical Universe.

\subsection{Fig. 2.}

A sketch of the classical phase space. Only half is shown, the other half
being the mirror image. We can see the stable (elliptic) fixed point at the
origin, and one of the unstable (hyperbolic) fixed points. The separatrices
connect the unstable points to each other, and divide the region of
periodical motion (within) from the region of unbound motion (outside). The
normalization is $b^{\prime }=b/2\sqrt{E_s}$, $p^{\prime }=p/\sqrt{2E_s}$.

\subsection{Fig 3.}

A qualitative plot of the equilibrium distribution function, as a function
of $E/E_s$. We can appreciate the divergence towards the origin, the
exponential rise towards the separatrix, and the falling off in the
inflationary region. For generic values of the cutoff, the area under the
curve is dominated by the peak at the separatrix.

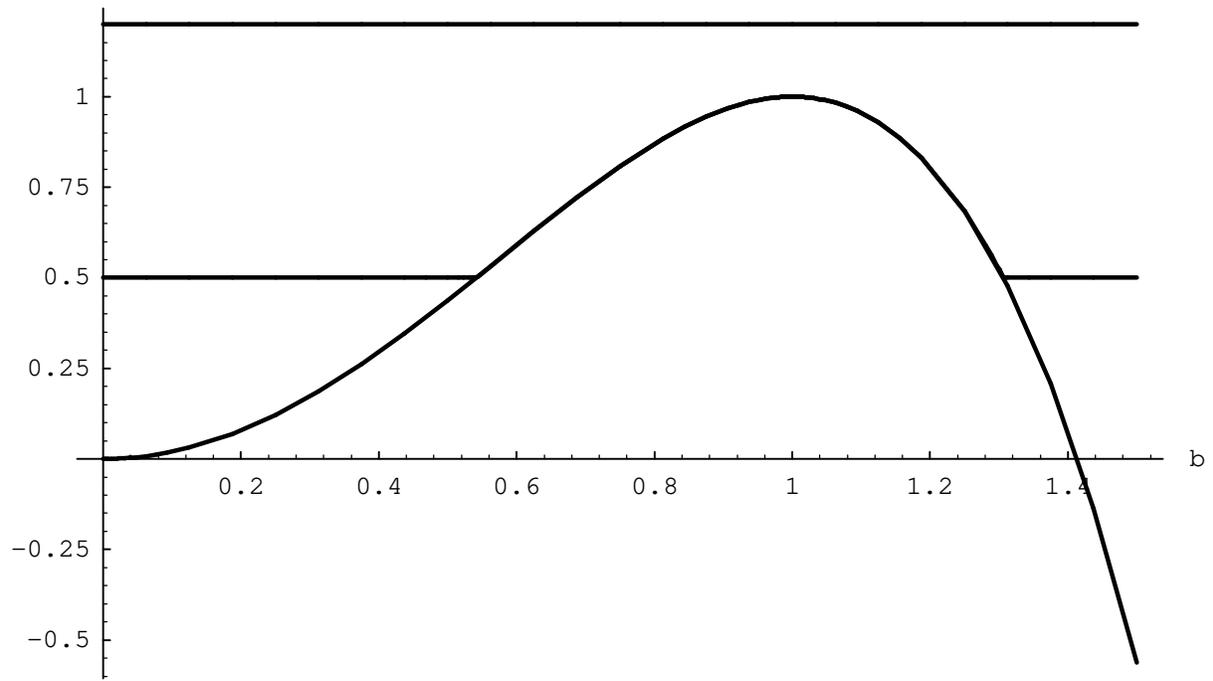
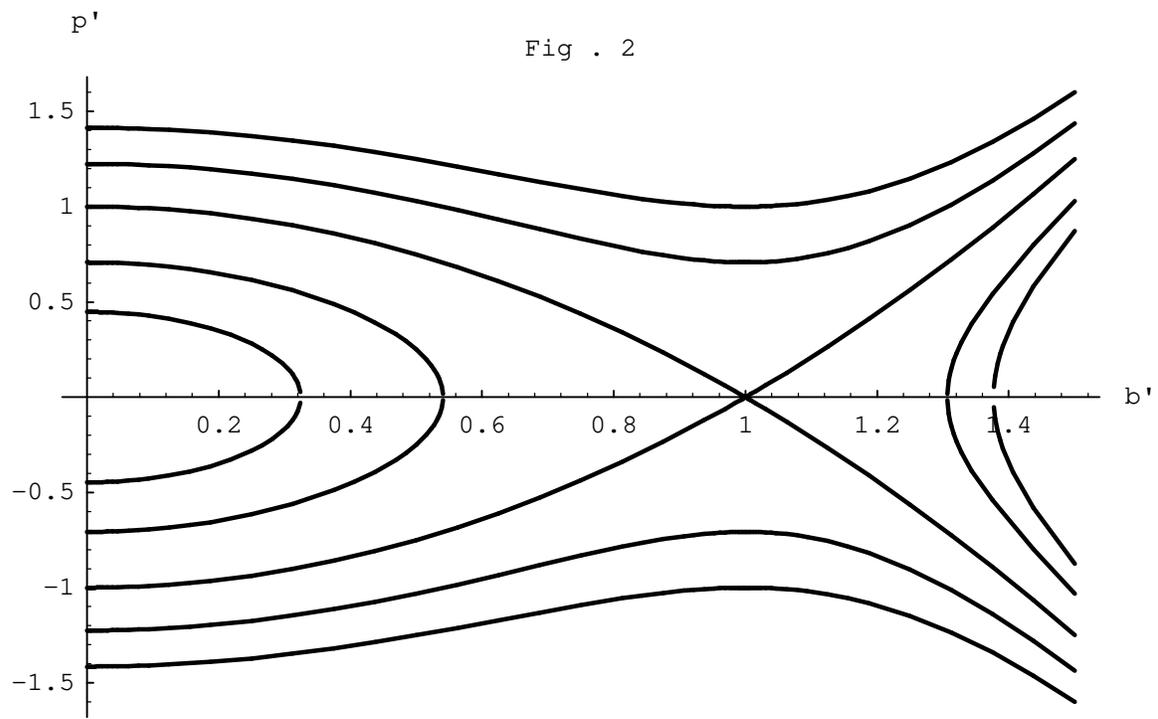
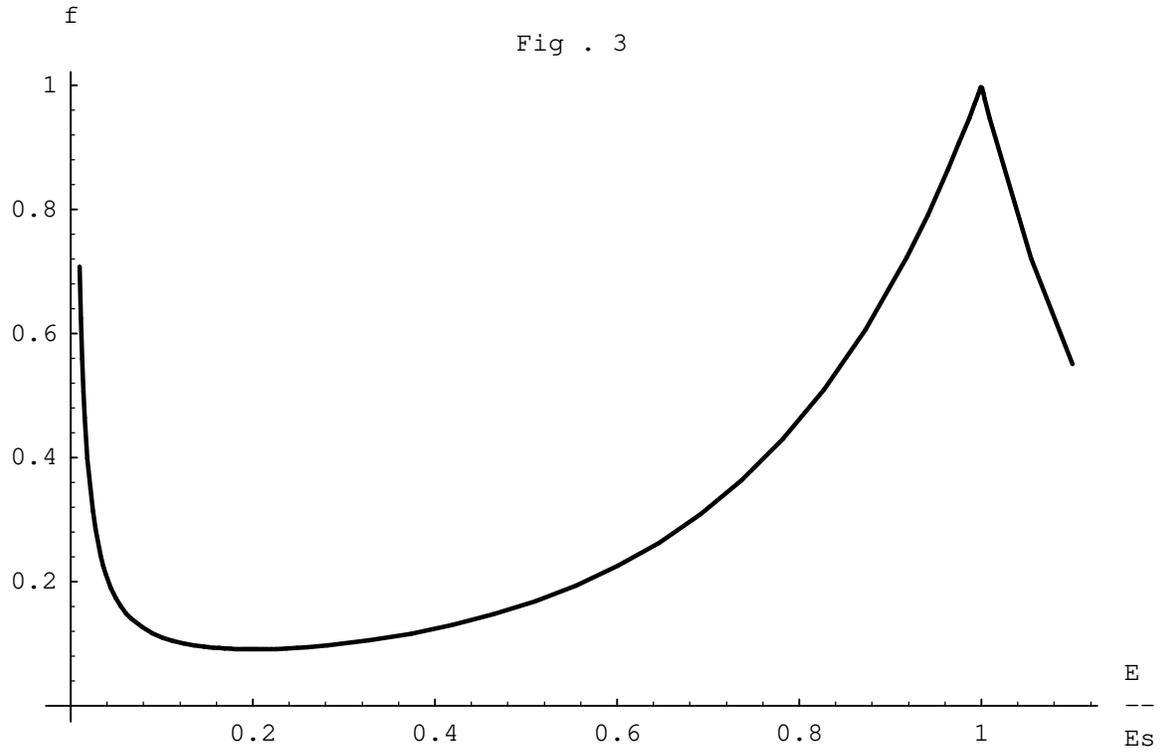
\end{document}